\documentclass[12pt,a4paper]{article}

\usepackage{times}
\usepackage{jheppub}
\makeatletter
\def\@fpheader{\relax}
\makeatother

\usepackage{amsmath}
\usepackage{amssymb}
\usepackage{hyperref}
\usepackage{braket}
\usepackage[T1]{fontenc}
\usepackage{tikz,pgf}
\usetikzlibrary{shapes}
\usetikzlibrary{calc}
\usetikzlibrary{decorations.pathmorphing}
\usetikzlibrary{decorations.pathreplacing,shapes.misc}
\usetikzlibrary{positioning}
\usetikzlibrary{arrows}
\usetikzlibrary{decorations.markings}

\tikzset{snake it/.style={decorate,decoration={snake,segment length=1.5mm, amplitude=.3mm}}}
\tikzset{biggerarrow/.style={
    decoration={markings,mark=at position 1 with {\arrow[scale=1.5]{>}}},
    postaction={decorate},
    shorten >=0.4pt}}
\tikzset{arrow at middle/.style={decoration={
    markings,
    mark=at position 0.5 with {\arrow{>}}}}}

\newcommand{\bC}{\mathbb{C}}

\def\cH{\mathcal{H}}
\def\cO{\mathcal{O}}
\def\tr{\mathop{\mathrm{tr}}\nolimits}
\def\kket#1{|#1\rangle\!\rangle}

\begin{document}

\title{Physics at the entangling surface}
\author[1]{Kantaro Ohmori}
\author[1,2]{and Yuji Tachikawa}
\affiliation[1]{Department of Physics, Faculty of Science, \\
 University of Tokyo,  Bunkyo-ku, Tokyo 133-0022, Japan}
\affiliation[2]{Kavli IPMU, \\
 University of Tokyo,  Kashiwa, Chiba 277-8583, Japan}
 \preprint{IPMU-14-0131, UT-14-28}

 \abstract{
         To consider the entanglement between the spatial region $A$ and its complement in a QFT,
         we need to assign a Hilbert space $\mathcal{H}_A$ to the region.
         Usually, some  boundary condition on $\partial A$ is implicitly chosen, but 
         we argue  that the choice of the boundary condition at $\partial A$ is physically meaningful and affects the subleading contributions to the entanglement R\'enyi entropy.
         We investigate these issues in the context of 2d CFTs, and show that we can indeed read off the Cardy states of the $c=1/2$ minimal model from the entanglement entropy of the critical Ising chain. 
}

\maketitle

\section{Introduction}

To define the entanglement of a quantum state $\ket{\psi}$, we write  the total Hilbert space $\cH$ as  the tensor product of the form  \begin{equation}
\cH=\cH_A\otimes \cH_B\label{naive}.
\end{equation} The reduced density matrix $\rho_A$ for $\cH_A$ is given by \begin{equation}
\rho_A=\tr_{\cH_B}  \ket{\psi}\bra{\psi}. \label{naivereduced}
\end{equation} The entanglement R\'enyi entropy is then \begin{equation}
S_n = (1-n)^{-1}\log\tr \rho_A^n \label{renyidef}
\end{equation} where $n$ can be non-integer, and the standard entanglement entropy is its $n\to 1$ limit.

Both in the lattice-based models (such as spin systems and lattice gauge theories) and  in the continuum quantum field theories, we usually want to measure the entanglement between the degrees of freedom associated to a spatial region $A$ and its complement $B$.
To have a decomposition of the total Hilbert space as in Eq.~\eqref{naive}, 
one needs to specify a boundary condition at the entangling surface, i.e.~at the boundary between the regions $A$ and $B$.
Often, the choice of such a boundary condition is done implicitly. 
In some cases, there is a very natural boundary condition to be used here. 
For example, in quantum spin systems, we can take $\cH_A$ and $\cH_B$ to be given by the tensor products of Hilbert spaces associated to sites contained in $A$ and in $B$.  This can be called a ``clear-cut'' decomposition of the spin system.

However, in more general cases, the choice of the boundary conditions is not a trivial matter.
For example, consider lattice gauge theories, that have degrees of freedom on each link on a lattice.  
It is known that there is no tensor product decomposition in lattice gauge theories due to the requirement of the gauge invariance, and the effect has been discussed in e.g.~\cite{Donnelly:2011hn,Casini:2013rba,Radicevic:2014kqa}.

As another example, consider any QFT. If it is weakly coupled, one might implicitly put the boundary condition by saying that the scalar field or the fermion field satisfies the Dirichlet or the Neumann boundary conditions. But in a strongly coupled theory, there is no natural choice of the boundary conditions. 
Take any strongly-coupled 2d conformal field theory (CFT). To talk about the Hilbert space associated to the region $A$, one needs to pick a boundary condition at $\partial A$.  Natural ones are the Cardy boundary conditions. One might say that the Cardy boundary condition associated to the unit operator would be the most natural one, but as we will see later, in the critical Ising model, the standard ``clear-cut'' decomposition does correspond to a Cardy boundary condition, but it is not the one associated to the unit operator. 

Our main point in this paper is that this choice of the boundary condition at the entangling surface can have interesting physical effects on the entanglement entropy, contrary to an often-found remark  that such a choice just changes the non-universal part of the entanglement entropy. 

\begin{figure}
\centering
	\begin{tikzpicture}[ultra thick,scale=.8]
		\draw (0,0) coordinate (a) -- ++(2,0) coordinate (b) ++ (1,0) coordinate (c)
		-- ++(2,0) coordinate (d) ++(1,0) coordinate (e) -- ++(2,0) coordinate(f);
		\draw[orange,snake it] (b) arc(180:360:.5 and .25) ;
		\draw[orange,snake it] (d) arc(180:360:.5 and .25) ;
		\draw ($(a) + (0,-1)$) coordinate (g) -- ($(f)+(0,-1)$) coordinate (h);
%		\node[anchor= east] at (a) {$ t=0 $};
%		\node[anchor= east] at (g) {$ t=-T $};
		\node[anchor= west] at (f) {$ \iota\ket{\psi}\in \mathcal{H}_{A,a} \otimes \mathcal{H}_{B,\bar{a}} $};
		\node[anchor= west] at (h) {$ \ket{\psi}\in \mathcal{H} $};
		\node[anchor= south] at ($(a)!0.5!(b)$) {B};
		\node[anchor= south] at ($(c)!0.5!(d)$) {A};
		\node[anchor= south] at ($(e)!0.5!(f)$) {B};
		\draw[|->, thick] ($(h)+(.4,.3)$) -- ($(f)+(.4,-.3)$);
		\node at ($(b)!0.5!(c)+(0,-.5)$) {$a$};
		\node at ($(d)!0.5!(e)+(0,-.5)$) {$a$};
		\draw[|<->|,thick] ($(b) + (0,.3)$) coordinate (ab) -- ($(c) + (0,.3)$) coordinate (ac);
		\node[anchor=south] at ($(ab) !.5! (ac)$) {$\epsilon$};
		\draw[|<->|,thick] ($(d) + (0,.3)$) coordinate (ad) -- ($(e) + (0,.3)$) coordinate (ae);
		\node[anchor=south] at ($(ad) !.5! (ae)$) {$\epsilon$};
		\draw[|<->|,thick] ($(c) + (0,-.3)$) coordinate (bc) -- ($(d) + (0,-.3)$) coordinate (bd);
		\node[anchor=north] at ($(bc) !.5! (bd)$) {$L$};
	\end{tikzpicture}
	\caption{ The cutting operation is given by a linear map $\iota :\mathcal{H}\to\mathcal{H}_{A,a} \otimes \mathcal{H}_{B,a}$. 
		The wiggly line represents ``thickened'' entangling surface with a boundary condition $a$ specified. 
		\label{fig:cut}}
\end{figure}
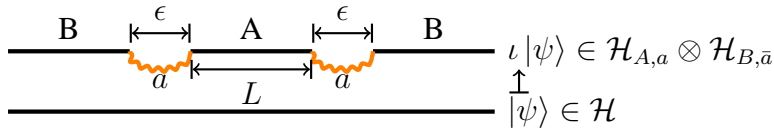

To discuss the effects of the boundary conditions at the entangling surface, it is helpful to have a formalism that makes manifest how the boundary conditions enter the definition of the entanglement entropy. 
So, pick a boundary condition $a$ at the entangling surface $\partial A$ between the regions $A$ and $B$. 
Now, we \emph{physically cut} the total space into $A$ and $B$ by inserting a physical boundary of thickness $\epsilon$ between the two regions, see Fig.~\ref{fig:cut}. 
Note that this operation applies both in the QFTs in the continuum  and in the quantum spin/lattice systems. 

Then we have the Hilbert spaces $\cH_{A,a}$, $\cH_{B,a}$ of the QFT on $A$ and $B$ with the boundary condition $a$ imposed on $\partial A$. 
The Hilbert spaces $\cH$ and $\cH_{A,a}\otimes\cH_{B,a}$ can be related by path-integration shown in the Fig.~\ref{fig:cut}.
We represent this cutting operation  by a linear map \begin{equation}
\iota: \cH \to \cH_{A,a} \otimes\cH_{B, a}.
\end{equation}

Given a state $\ket{\psi}\in \cH$,  the state after the cut is $\iota\ket{\psi}$ and lives in the tensor product. We can now form the reduced density matrix associated to the region $A$ as follows: \begin{equation}
\rho_A= \tr_{\cH_{B, a} }  \iota \ket{\psi}\bra{\psi} \iota^\dagger.
\end{equation}
The   entanglement entropy with respect to the cutting operation $\iota$ using the boundary condition $a$  is then defined by the formula \eqref{renyidef}.

\paragraph{Comments on our description of the entanglement entropy:}
Before getting into the discussions of concrete examples, 
we give some comments of our way of writing down the entanglement entropy, with the dependence on the boundary conditions made manifest. 

\begin{itemize}
\item  
In certain very simple systems such as the quantum spin systems, there is a natural ``clear-cut'' boundary condition. Then our way of writing down the entanglement entropy reduces to the usual one, by taking $\epsilon=0$ and $\iota=1$. 

Note that even in this relatively clear-cut case, there are two issues. First, whether the Hilbert space of the system is clearly cut depends on the choice of the main observables. When we perform a non-local redefinition such as  the Jordan-Wigner transformation, a decomposition that is clear-cut to one set of observables might be not clear-cut to another set of observables. 
Second, the coarse-graining to the infrared should be carefully taken. 
In order not to destroy the tensor product decomposition of the wavefunction under renormalization, we need to modify the Hamiltonian acting on the time slice where $\ket\psi$ lives so that there is no cross-talk between $H_A$ acting on $\mathcal{H}_A$ and $H_B$ acting on $\mathcal{H}_B$. The sum $H_A+H_B$ would then be different from $H_{A+B}$ on the entangling surface.
Then, renormalization with such modification makes $\iota$ not to be an identity in the infrared continuum QFT.
The resulting boundary condition and the cutting operation can also depend on the choice of the ultraviolet realizations, if there are multiple such realizations that flow to the same infrared QFT.  

\item In more generic systems, to say what the Hilbert spaces $\cH_{A,B}$ associated to the subregions $A$, $B$ are, we need to specify the boundary condition at the entangling surface. 
What we are doing here is just making the boundary condition explicit, whereas it is usually made implicit in the literature. 
For example, in the literature of the entanglement entropy of the quantum field theory, the entanglement entropy is often implicitly defined by the replica method, without explicitly specifying the boundary condition of the theory at the entangling surface.  Here we are making manifest the choice of the boundary condition at the singularity of the replica manifold. 

\item Our way of writing down the entanglement entropy might be compared to the following gedanken experimental situation: even in a quantum spin system, to form a density matrix of a subsystem, one would need to physically separate the region $A$ and the region $B$ by performing an operation, which is represented by $\iota$.  Afterwards, one destroys (or averages over) the conditions in the region $B$, to form the density matrix $\rho_A$. We consider the von Neumann entropy of this $\rho_A$, which is our working definition of the entanglement entropy in a general situation. 

\item In the previous analysis of the entanglement entropy of the lattice gauge theories in e.g.~\cite{Donnelly:2011hn,Casini:2013rba,Radicevic:2014kqa}, one of the central issues was how to deal with the Hilbert space associated to the links that intersect with the entangling surface. Many methods to deal with them were proposed. Our point of view is that we can be agnostic about what is the best method; various methods just define various different $\iota$. Note also that by treating the links that intersect with the entangling surface specially, they are effectively putting boundary conditions there. 

As can be seen from this example of lattice gauge theories, the map $\iota$ is \textit{not} unique nor canonical in any way. It depends not only on boundary conditions but also the shape of the ``thickend'' entangling surfaces and the distance between the input and output time slices of Fig.~\ref{fig:cut}. 
$\iota$ is just an operation which relates total Hilbert space $\mathcal{H}$ and tensor product space 
$\cH_{A,a} \otimes\cH_{B, a}$.

\end{itemize}

\paragraph{Organization:}
In the rest of the paper, we first compute the entanglement entropy defined in this manner for a single segment of length $L$ of any 2d CFT, and show the large $L$ behavior \begin{equation}
S_n \sim  (1+\frac{1}{n})\frac{c_\text{eff}}{6} \log\frac{L}{\epsilon}  + c' + c'' (\frac{L}{\epsilon})^{-2\Delta/n}+ \cdots\label{mainresult}.
\end{equation} 

The leading logarithmic piece reproduces the known universal result \cite{Holzhey:1994we,Calabrese:2004eu,Bianchini:2014uta}. 
Here, $c_\text{eff}$ is the so-called effective central charge characterizing the asymptotic density of states of the 2d CFT. 
The constant term $c'$ now depends on the boundary condition, and can be written in terms of the boundary entropy, introduced in \cite{AffleckLudwig}.  %Note that the constant term is now meaningful, since the UV cut-off $\epsilon$ in the logarithmic term is defined as the physical thickness of the regularized entangling surface. 

The subleading correction of the form $c''(L/\epsilon)^{-2\Delta/n}$ was already discussed \cite{Cardy:2010zs}; in our approach, we clearly see which $\Delta$ appears as the exponent, and how it depends on the boundary condition.
Such coefficients of subleading corrections are thought to be ``non-universal'', but in the next section we will see we can extract ``semi-universal'' information from such coefficients from a certain limit.

Next, we confirm these considerations by a study of the critical Ising model, which flows to the $c=1/2$ minimal model. 
In the ultraviolet, there is a clear-cut decomposition as in \eqref{naive}, and the entanglement entropy of a single segment was studied in \cite{Vidal:2003,Jin:2004aaa,Calabrese:2010aab}.  
By extending their analysis, we show that there is indeed a Cardy boundary condition $\ket{\sigma}$ inserted at the entangling point. 
%In this case $\Delta$ in Eq.~\eqref{mainresult} is $1$.

The $c=1/2$ minimal model also has two other Cardy boundary conditions $\ket{1}$ and $\ket{\varepsilon}$. We show that they correspond in the ultraviolet to a cutting operation defined as follows. We first split the total Hilbert space as $\cH_\text{tot} = \cH_L \otimes \bC^2 \otimes \cH_R$. Then we measure the spin at the central site. Then the partial trace is taken over $\cH_R$, giving a density matrix on $\cH_L$. 
We will see that the entanglement entropy computed with this cutting operation  reproduces the Cardy conditions $\ket{1}$ and $\ket{\varepsilon}$ as ''semi-univeral" information stated above. %In this case $\Delta$ in Eq.~\eqref{mainresult} is $1/8$.
 
\section{2d CFT analysis}\label{sec:2dCFT}
First, let us calculate the entanglement entropy of the single segment for the vacuum state $\ket{\Omega}$ of  two-dimensional CFTs.   The computation basically follows the one  in  \cite{Holzhey:1994we}, except that we have boundary conditions at the entangling surface.
Note that in two dimensions, the entangling surface consists of two points,
and we can put two different boundary conditions $a_1$, $a_2$ there. 
Therefore, we use a cutting operation of the form \begin{equation}
\iota: \cH\to \cH_{a_1,A,a_2} \otimes \cH_{a_2,B,a_1}.
\end{equation}
We do not assume that $a_{1,2}$ are conformal boundary conditions. 

\subsection{Setup}

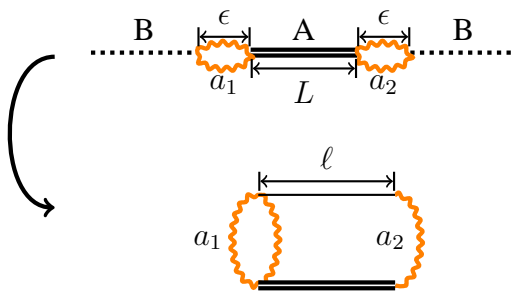
\begin{figure}
\centering
\begin{tikzpicture}
\node (X) {\begin{tikzpicture}[ultra thick,scale=.7]
		\draw[dotted] (0,0) coordinate (a) -- ++(2,0) coordinate (b) ++ (1,0) coordinate (c)
		 ++(2,0) coordinate (d) ++(1,0) coordinate (e) -- ++(2,0) coordinate(f);
		\draw[double] (c)--(d);
		\draw[orange,snake it] (b) arc(0:360:-.5 and -.25) ;
		\draw[orange,snake it] (d) arc(0:360:-.5 and -.25) ;
%		\node[anchor= west] at (f) {$ \iota\ket{\Omega}\bra{\Omega} \in \cH_{a_1,A,a_2}\otimes \cH_{a_1,A,a_2}{}^\dagger$};
		\node[anchor= south] at ($(a)!0.5!(b)$) {B};
		\node[anchor= south] at ($(c)!0.5!(d)$) {A};
		\node[anchor= south] at ($(e)!0.5!(f)$) {B};
		\node at ($(b)!0.5!(c)+(0,-.6)$) {$a_1$};
		\node at ($(d)!0.5!(e)+(0,-.6)$) {$a_2$};
		\draw[|<->|,thick] ($(b) + (0,.3)$) coordinate (ab) -- ($(c) + (0,.3)$) coordinate (ac);
		\node[anchor=south] at ($(ab) !.5! (ac)$) {$\epsilon$};
		\draw[|<->|,thick] ($(d) + (0,.3)$) coordinate (ad) -- ($(e) + (0,.3)$) coordinate (ae);
		\node[anchor=south] at ($(ad) !.5! (ae)$) {$\epsilon$};
		\draw[|<->|,thick] ($(c) + (0,-.3)$) coordinate (bc) -- ($(d) + (0,-.3)$) coordinate (bd);
		\node[anchor=north] at ($(bc) !.5! (bd)$) {$L$};
	\end{tikzpicture}};
\node[anchor=north] (Y) at (X.south)  {	\begin{tikzpicture}[scale = 0.6,ultra thick, baseline =-20]
		\draw[orange,snake it] (0,0) coordinate(a) arc (90:450:.5 and 1);
		\draw[thick] (a) -- ++ (3,0) coordinate (b);
		\draw[orange,snake it] (b) arc (90:-90:.5 and 1) coordinate (c);
		\draw[double] (c) -- ++ (-3,0);
		\node[anchor=east] at (-0.5,-1) {$a_1$};
		\node[anchor=east] at (3.5,-1) {$a_2$};
		\draw[thick,|<->|] ($(a)+(0,.3)$) -- ($(b)+(0,.3)$);
		\node[anchor=south] at ($(a)!.5!(b)+(0,.3)$) {$\ell$};
	\end{tikzpicture}};
\draw[ultra thick,->,out=180,in=180] ($(X.west)+(-.3,0)$) to ++ (0,-2);
\end{tikzpicture}
	\caption{ The path-integral expression for the reduced density matrix $\rho_A$.   
		For two-dimensional QFTs, the entangling surface consists of two points, where different conditions $a_1$, $a_2$ can be specified. 
		\label{fig:2cut}}
\end{figure}

The vacuum state $\ket{\Omega}\in \cH$ at the time $t=0$ can be prepared by the path integral over the half-plane $t<0$. The state after the cutting, $\iota\ket{\Omega}$, can be implemented by inserting two very small physical boundaries of size $\epsilon$ at the two entangling points, with boundary conditions $a_1$ and $a_2$ specified. Similarly, the bra $\bra{\Omega}\iota^\dagger$ can be prepared by the path integral over the half-plane $t>0$. The partial trace over $\cH_{a_2,B,a_1}$ is then done by the path integral over the region $B$. 
At the end, we see that the reduced density matrix $\rho_A$ on $\cH_{a_1,A,a_2}$ is represented by the geometry shown on the top of Fig.~\ref{fig:2cut}, where we perform the path integral over the whole plane with two boundaries of size $\epsilon$ at the entangling points, with a slit on the segment $A$. 

Using the conformal transformation, the same path integral can be calculated by the path integral over the cylinder with circumference $2\pi$ and width $\ell$, up to a factor given by the anomalous Weyl transformation. For example, one can chose the conformal transformation expressed as $z\mapsto \log z-\log (L-z)$. The resulting geometry may be not a cylinder with flat boundaries depending on the precise shape of the ``thickened entangling points'', but one can always flatten the boundaries with a further small conformal transformation. 
 The expression of $\ell$ in terms of $L$ and $\epsilon$ is determined by the particular choice  of the shape of the thickened entangling point. Here we adopt $\ell= \log (L/\epsilon)^2 +\mathcal{O}((L/\epsilon)^{-1})$ as the precise definition of $\epsilon$. 
 The $\mathcal{O}((L/\epsilon)^{-1})$ part depends on the precise shape of the ``thickened entangling points'', but we ignore such terms through this paper.
Also, it is to be noted that such conformal transformation would affect the boundary conditions $a_{1,2}$ in general if they were not conformal to start with. Not to overcomplicate the notations, from now on, we use the symbol $a_{1,2}$ to denote the boundary conditions on the two ends of the cylinder after these conformal transformations are made. 
 
Now we see that 
\begin{equation}
\tr \rho_A^n = {Z_n}/{Z_1^n} \label{trn}
\end{equation} 
where $Z_n$ is the cylinder partition function with the boundary conditions $a_{1,2}$, 
such that the circumference is $2\pi n$ but the width is still the same $\ell\sim \log (L/\epsilon)^2$.
We can perform a conformal transformation to make the circumference $2\pi$ and the width $\ell/n$.
This makes the boundary conditions coarse-grained by a scale factor of $n$.
Let us denote the resulting boundary conditions by $a_{1,2}^{(n)}$. We have 
\begin{equation}
	Z_n=\braket{a_1^{(n)}| \exp\left(\frac{\ell}{n} (\frac{c+\bar{c}}{24}- L_0-\bar{L}_0)\right) |a_2^{(n)}} 
	\label{eq:Zn}
\end{equation}
where  $\ket {a_{1,2}^{(n)}}$ are the corresponding boundary states.
The boundary condition $\ket{a_i^{(n)}}$ will flow in the limit $n\to \infty$ to some conformal boundary condition. If we insert conformal boundary conditions $a_i$ at the entangling points from the start, $a_i^{(n)}$ is independent of $n$,  $a_i^{(n)}\equiv a_i$.

\subsection{Leading universal piece}

Let $\ket 0$ be the state with lowest dimension that couples to both boundary states $\ket{a_1}$ and $\ket{a_2}$. For 2d unitary CFTs with discrete spectrum, $\ket 0$ is the vacuum $\ket{\Omega}$, but this can be in general different in CFTs with continuous spectrum or non-unitary theories.
For example, in the Liouville theory with paramter $Q$ such that $c=1+6Q^2$, the dimension $\Delta_0$, which is the sum of the chiral and the anti-chiral dimension, of the operator corresponding to the state $\ket0$ is given by $Q^2/2$. Note that for non-chiral operators, the dimension is the sum of $L_0$ and $\bar L_0$.
In the limit $L/\epsilon \gg 0$, the leading piece of $Z_n$ can  easily be evaluated: 
\begin{equation}
	Z_n 
	\sim \braket{a_1^{(n)}|0} \exp\left(\frac{\ell}{n}(\frac{c}{12}-\Delta_0)\right) \braket{0|a_2^{(n)}}.
\end{equation} Plugging this into Eq.~\eqref{trn} and using $\ell\sim\log(L/\epsilon)^2$, we find \begin{equation}
S_n \sim (1+\frac1n) \frac{c_\text{eff}}{6} \log\frac{L}{\epsilon} +\cdots 
\end{equation} where $c_\text{eff}=c-12\Delta_0$ is the effective central charge controling the high-temperature behavior of the cylinder or torus partition function. When $\ket0$ is the vacuum, $\Delta_0=0$ and $c_\text{eff}=c$. For the Liouville theory, however, $c_\text{eff}=1$ independent of the paramter $Q$. That $c_\text{eff}$ appears as the coefficient of the logarithmic term was first pointed out in \cite{Bianchini:2014uta}.

\subsection{Subleading corrections}

Let us restrict our attention to the 2d unitary CFTs whose primary states are discrete.
Let $\ket\cO$ be the state with the second-lowest dimension that couples  to both boundary states $\ket{a_1}$ and $\ket{a_2}$.  Denote its dimension by $\Delta_\cO$.
Note that by assumption, the lowest state is the vacuum $\ket0=\ket\Omega$. 
We have
\begin{multline}
	Z_n = 
	\braket{a_1^{(n)}|0} \exp\left(\frac{\ell}{n}\frac{c}{12}\right) \braket{0|a_2^{(n)}}  \\
	+ \braket{a_1^{(n)}|\mathcal{O}}\exp\left(\frac{\ell}{n}(\frac{c}{12}-\Delta_\cO)\right) \braket{\mathcal{O}|a_2^{(n)}} +\cdots.
\end{multline} 
Then the R\'enyi entropy $S_n = \log (\tr{\rho_A^n})/(1-n)$  has the expansion
\begin{multline}
	S_n= (1+\frac1n)\frac{c}{6}(\log\frac{L}{\epsilon})  \\
	+ \frac{1}{1-n}(s(a_1^{(n)})-ns(a_1^{(1)}) + s(a_2^{(n)})-ns(a_2^{(1)})) \\
	+ \frac{1}{1-n}
	\frac{\braket{a_1^{(n)}|\mathcal{O}}\braket{\mathcal{O}|a_2^{(n)}}}{\braket{a_1^{(n)}|0}\braket{0|a_2^{(n)}}}
	\left(
	\frac{L}{\epsilon}
	\right)^{-2\Delta_{\mathcal{O}}/n}
	+\cdots
	\label{eq:nonconfRenyi}
\end{multline}   
Here,
 the quantity $s(a) = \log \braket{a|0}$ for a boundary condition $a$ is the Affleck-Ludwig
boundary entropy.
The corrections  to \eqref{eq:nonconfRenyi} come either from the contribution to $Z_n$ of the next operator $\mathcal{O'}$ that couples to the boundary states or from the contribution to $Z_1$ from $\mathcal{O}$ itself, and therefore is of the order  $O(e^{-\ell \min(\Delta_{\mathcal{O}},\Delta_{\mathcal{O'}}/n)})$ .

We already discussed the leading logarithmic piece  independently of the boundary condition.
The constant term, which is now meaningful because the cutoff $\epsilon$ is just the size of the boundary inserted at the entangling point, is determined by the boundary entropies. 

The subleading correction of the form $(L/\epsilon)^{-2\Delta_\cO/n}$ to the R\'enyi entropy as in Eq.~\eqref{eq:nonconfRenyi}, was already discussed in \cite{Cardy:2010zs}. 
We now understand which operator $\cO$ gives rise to this subleading correction: 
it is the operator with lowest dimension (apart from the identity) that has non-zero overlap with both the boundary conditions, i.e.~$\braket{a_1|\cO}$ and $\braket{a_2|\cO}$ are both nonzero.
Therefore, the operator $\cO$ and the dimension $\Delta_\cO$ appearing in the subleading correction can and do depend on the choice of the boundary conditions $a_{1,2}$.
Such dependence on the boundary conditions is noticed in \cite{2010JSMTE..09..003C} in the setting of the entanglement entropy of the slightly off-critical systems computed in terms of the corner transfer matrix. Here, we directly see its origin in the 2d CFT language.

As $a_i^{(n)}$ is obtained by a coarse-graining of $a_i$ by a scale factor of $n$,  the $n\to\infty$ limit $a_i^{(\infty)}$ should become  conformal boundary conditions. 
Thus, we obtain the following result for a certain scaling limit of entanglement R\'enyi entropy with a fixed parameter $q<1$ as follows.
First, we can get the following result for the partition function $Z_n$:
\begin{align}
	\lim_{n\to\infty}Z_n|_{\frac{L}{\epsilon}=q^{-n/2}}=Z^\text{conf}(q;a_1^{(\infty)},a_2^{(\infty)}),
\end{align}
where $Z^\text{conf}(q;a_1^{(\infty)},a_2^{(\infty)}) =\bra{a_1^{(\infty)}} q^{L_0+\bar{L}_0-c/12}\ket{a_2^{(\infty)}}$ is the conformal cylinder partition function with modular parameter $q$ and conformal boundary conditions $a_1^{(\infty)}$ and $a_2^{(\infty)}$.
Because $Z_1$ can only have terms with powers independent of $n$, all the subleading contribution of the $S_n$ with power proportional to $1/n$ of $L/\epsilon$ should come from $Z_n$.
Therefore, if we define the fractional power part $S_n^\text{frac}$ as 
\begin{multline}
	S_n=(1+\frac{1}{n})\frac{c_\text{eff}}{6}\log\frac{L}{\epsilon}+\mathcal{O}(L^0)
	+S_n^{\text{frac}}
	+\text{(terms with powers independent of $n$)}.
\end{multline}
we get the following convergent limit:
\begin{multline}
	\lim_{n\to\infty} (1-n)S_n^\text{frac}|_{L/\epsilon =q^{-n/2}}=\\
	\log\left(q^{\frac{c_\text{eff}}{6}}Z^\text{conf}(q;a_1^{(\infty)},a_2^{(\infty)})\right)-s(a_1^{(\infty)})-s(a_2^{(\infty)}).
	\label{eq:scaling}
\end{multline}
The boundary entropies $s(a_i^{(\infty)})$ can also be obtained from $L^0$ part of $S_n$ as is expressed in \eqref{eq:nonconfRenyi}, therefore we can construct all-order cylinder partition functions from the asymptotic behavior of $S_n$.
Note that this result do not require that the original boundary conditions $a_1$ and $a_2$ are conformal.
Thus, if we assume that the UV lattice model calculation should correspond to a continuum calculation with some (not necessarily conformal) boundary conditions, 
we should be able to obtain conformal boundary conditions by taking the limit described above. 
The result depend only on what conformal boundary conditions $a_i$ flow to,
therefore such dependence can be called ``semi-universal".
Let us see this actually happens in the case of the critical Ising chain.

\section{Critical Ising model}\label{sec:2dising}

The critical Ising model has the Hamiltonian \begin{equation}
H= \sum_{i=-\infty} ^\infty (\sigma^z_i\sigma^z_{i+1} + \sigma^x_i).
\end{equation} 
We introduce Majorana fermion operators $\gamma_{P,i}$ and $\gamma_{Q,i}$ 
with the anticommutator $\{\gamma_{\mu,i},\gamma_{\nu,j} \}=2\delta_{\alpha\beta}\delta_{ij}$ where $\mu,\nu=P,Q$.
Under the transformation 
\begin{align}
\sigma^x_i &= \mathrm{i}\gamma_{P,i} \gamma_{Q,i}, &
\sigma^z_i &= -\gamma_{Q,i} \prod_{j=i+1}^\infty \mathrm{i}\:\gamma_{P,j} \gamma_{Q,j}
\end{align} the Hamiltonian becomes 
\begin{equation}
H= \sum_{i=-\infty} ^\infty \mathrm{i}\:(\gamma_{Q,i-1}\gamma_{P,i} + \gamma_{P,i}\gamma_{Q,i}).
\end{equation} Here and in the following we use $\mathrm{i}$ to denote the imaginary unit.
Non-zero 2-point functions of the fermion operators are (see e.g.~\cite{Vidal:2003})
\begin{align}
	\braket{\gamma_{P,i}\gamma_{P,i}}&=	\braket{\gamma_{Q,i}\gamma_{Q,i}}=1,\\
	\braket{\gamma_{P,i}\gamma_{Q,j}}&=\frac{-2\mathrm{i}}{\pi(2(i-j)+1)}.
	\label{eq:2pt}
\end{align}

\subsection{Free boundary condition}
We first review the computation of the entanglement R\'enyi entropies $S_n$ with the clear-cut decomposition \eqref{naive}, done in \cite{Jin:2004aaa,Calabrese:2010aab}. What we add here is that the Cardy state $\ket{\sigma}$ can be reproduced in its $n\to \infty$ limit. 

Take the fermion operators on the sites from $i=1$ to $i=L$ to be the region $A$, and all the other sites to be the region $B$.
Then we have the clear-cut decomposition of the Hilbert space as in Eq.~\eqref{naive}. 
Note that this decomposition is not identical to the clear-cutting with respect to spin operators.
The reduced density matrix $\rho_{cc,A}$ (where $cc$ stands for `clear-cut') can be reconstructed just from the fact that the Majorana fermion operators $\gamma_{P,i}$, $\gamma_{Q,i}$ for $i=0,\ldots,L$ have the 2-point functions \eqref{eq:2pt}. 
By  block-diagonalizing the 2-point function in terms of an $SO(2L)$ matrix, we can write the reduced density matrix as a collection of free fermion modes with occupation numbers given by $\nu_i$:
\begin{equation}
\rho_{cc,A} = \bigotimes_{i=0}^L \rho_i, \quad \rho_i=\textrm{diag}(\frac{1+\nu_i}2,\frac{1-\nu_i}2)
\label{eq:diagonal}
\end{equation} and we can compute the entropy from this.
In \cite{Vidal:2003} the von Neumann entropy was computed numerically and nicely fitted to $(c/3)\log L$ with $c=1/2$. This behavior was later proved in \cite{Jin:2004aaa}, using the fact that 
the matrix $\rho_{cc,A}$ is a Toeplitz matrix, whose asymptotic behavior of eigenvalues is known mathematically.  

We denote the characteristic polynomial of $\rho_{cc,A}$  by $D_{L}(\lambda) = \prod_m^{L} (\lambda^2-\nu_m^2)$.
In \cite{Calabrese:2010aab} the corresponding quantity $D_L^{XX}(\lambda)$ for the XX model
was considered, where $D_L(\lambda)^2=D_{2L}^{XX}(\lambda)$.
As explained there, the generalized Fisher-Hartwig conjecture states that
\begin{multline}
	D^{XX}_{L}(\lambda) \sim (\lambda+1)^{L/2}(\lambda-1)^{L/2}
	\sum_{m\in\mathbb{Z}} (2L)^{-2m-2\beta_\lambda} \mathrm{e}^{-\pi\mathrm{i} m L}\\
	\times \left( G(1+m+\beta_\lambda)G(1-m-\beta_\lambda)\right)^2,\label{gFH}
\end{multline}
where $\beta_\lambda=(2\pi\mathrm{i})^{-1} \log \frac{\lambda+1}{\lambda-1}$ and  $G$ is the Barnes G-function.
We choose the branch of the logarithm so that  $|\beta_\lambda|\le 1/2$. Note that the sum over $m$ is the sum over the branches. For the status of the generalized Fisher-Hartwig conjecture, see e.g.~\cite{DIKreview}.
We expect that the ratio of the left hand side and the right hand side is $1+O(L^{-1})$ in the large $L$ limit.

In terms of $D_L$, we can write the R\'enyi entropy as
\begin{equation}
	S_n(L)= \frac12\oint_C \frac{\mathrm{d}\lambda}{2\pi \mathrm{i}} s_n(\lambda) \frac{\mathrm{d}}{\mathrm{d} \lambda} \log D_{L}(\lambda).\label{contourintegral}
\end{equation}
where $s_n(\lambda)=\frac{1}{1-n}\log((\frac{1+\lambda}{2})^n+(\frac{1-\lambda}{2})^n)$ and $C$ is the contour encircling all poles of $\frac{\mathrm{d}}{\mathrm{d}\lambda}\log D_L(\lambda)$, which are on $[-1,1]$.
After some calculations detailed in \cite{Calabrese:2010aab}, we have
\begin{align}
	S_n(L)&= \frac{1+n}{12n}\log (4L) + \hat s_n + \check s_n + \mathcal{O}(L^{-1}),\\
	\hat s_n &= \frac{-2}{\pi}\int_{0}^\infty  s_n(\tanh\pi w)\Re \psi(\frac{1}{2}+\mathrm{i} w)\mathrm{d}w,\\
	\check s_n &=\frac{1}{1-n}\sum_{i=1}^{\infty}\log (1+(4L)^{-2(2i-1)/n}X_{\frac{2i-1}{2n}}),
	\label{eq:CalEss}
\end{align}
where $\psi$ is the digamma function and $X_w = (\Gamma(\frac{1}{2}+w)/\Gamma(\frac{1}{2}-w))^2$.
Here, the logarithmic term and $\hat s_n$ come from the $m=0$ term 
and $\check s_n$ come from the terms with $m=\pm1$  in Eq.~\eqref{gFH}.

Now we can extract the cylinder partition function
by keeping the terms of the form $L^{-\alpha/n}$  
in the asymptotic expansion in $L$.
First, we see that $\hat s_n=\hat s_\infty(1+1/n)+\mathcal{O}(1/n^2)$.
By comparing with Eq.~\eqref{eq:nonconfRenyi},
we conclude $s({a^{(\infty)}})=0$. Not only that, the scaling limit of \eqref{eq:scaling} becomes
\begin{equation}
	\lim_{n\to\infty}(1-n)S_n^\text{frac}|_{4L=q^{-n/2}}= \log\prod_{i=1}^\infty (1+q^{2i-1}),
\end{equation}
where we used $X_{(2i-1)/(2n)}\to 1$  when $n\to \infty$.
The right hand side indeed the right hand side of the \eqref{freefree} where boundary conditions $a_i^{(\infty)}$ correspond to the Cardy state $\ket\sigma$, as the conformal partition function is 
\begin{equation}
	Z^\text{conf}(q;\sigma,\sigma)= q^{-\frac{1}{24}}\prod_{i=1}^\infty (1+q^{2i-1}).
	\label{freefree}
\end{equation} 

To see this, recall the expansion of the Cardy states $\ket 1$, $\ket \varepsilon$ and $\ket \sigma$  in terms of the Ishibashi states $\kket 1$, $\kket\varepsilon$ and $\kket\sigma$:
\begin{equation}
\begin{aligned}
\ket 1&=2^{-1/2} \kket 1 + 2^{-1/2} \kket \varepsilon + {2^{-1/4}} \kket \sigma,\\
\ket \varepsilon&=2^{-1/2} \kket 1 + 2^{-1/2}\kket \varepsilon - {2^{-1/4}} \kket \sigma,\\
\ket \sigma&=\kket 1 - \kket \varepsilon.
\end{aligned}
\label{eq:boundary states}
\end{equation}
Then $\braket{\sigma|q^{L_0+\bar L_0-c/12}|\sigma}=\chi_1(q^2)+\chi_\varepsilon(q^2)$, which is exactly Eq.~\eqref{freefree}. Here $\chi_{1,\varepsilon,\sigma}(q)$ are the Virasoro minimal characters for the three primaries.

\subsection{Fixed boundary condition}

Next,  we calculate the entanglement entropy with the other cutting operations, involving observations of the spin at the entangling points.
Namely, we put the entangling points on top of the sites $i=0$ and $i=L$,
and introduce projection operators proportional to 
\begin{equation}
({1\pm \sigma_0^z})(1\pm \sigma_L^z).\label{eq:projection}
\end{equation}
We let the operators $\gamma_{P,i}$, $\gamma_{Q,i}$ from $i=1$ to $i=L-1$ together with $\gamma_{Q,0}$ and $\gamma_{P,L}$ to be the operators observable in the region $A$. 
The density of states after the projection is given  by \begin{equation}
\rho_{\pm\pm,A} \propto \rho_{cc,A} 
(1\pm \sigma_0^z)(1\pm \sigma_L^z)
\end{equation} 
up to normalization.

The expectation value of any observable $\mathcal{O}$ is then given by  $\tr \rho_{\pm\pm,A} \mathcal{O} /\tr \rho_{\pm\pm,A}$. 
The terms which include only one $\sigma^z$ does not contribute because $\sigma^z$ consists of infinitely many fermions.
The product $\sigma^z_0\sigma^z_L$ is 
\begin{equation}
	\sigma_0^z\sigma^z_L= \gamma_{Q,0} \left(\prod_{i=1}^{L-1}\mathrm{i}\gamma_{P,i}\gamma_{Q,i}\right)\gamma_{P,L}
\end{equation}
which is  the chirality operator $\Gamma$ of the $SO(2L)$ gamma matrices  $\gamma_{Q,0}$, \ldots, $\gamma_{P,L}$.
Therefore, in the basis where $\rho_{cc,A}=\bigotimes_i \rho_i$ as in Eq.~\eqref{eq:diagonal},
we have $\Gamma=\bigotimes_i \sigma^z_i $.

After normalization, we have 
\begin{equation}
	\tr \rho_{\pm\pm,A}^n = 2^{n-1}\frac{\tr\rho_{cc,A}^n \pm \tr\rho_{cc,A}^n \Gamma}{(1\pm\tr\rho_{cc,A}\Gamma)^n}
	\label{eq:fixedrho}
\end{equation} 
where the $\pm$ sign on the right hand side is the product of the $\pm$ sings on the left hand side, and \begin{equation}
\tr\rho_{cc,A}^n \Gamma=\prod_{i=0}^L \left(
(\frac{1+\nu_i}2)^n - (\frac{1-\nu_i}2)^n
\right).
\end{equation}

We can compute the asymptotic behavior of $U_n(L):=\log\tr\rho_{cc,A}^n\Gamma $ as before. We have \begin{equation}
U_n(L)
=\Re \frac12\oint_C \frac{\mathrm{d}\lambda}{2\pi \mathrm{i}}  u_n(\lambda+\mathrm{i}\epsilon) \frac{\mathrm{d}}{\mathrm{d} \lambda} \log D_{L}(\lambda) \label{ff}
\end{equation} where 
$u_n(\lambda)=\log((\frac{1+\lambda}{2})^n-(\frac{1+\lambda}{2})^n)$ and $C$ is the same contour as before. We added a small positive imaginary part $\mathrm{i}\delta$ in the argument of $u_n(\lambda)$ to deal with the branch cut in $u_n$, and we take $\delta \to +0$ later. 
At the end of the day, we have \begin{align}
	U_n(L)&= \frac{2-n^2}{12n}\log (4L) + \hat u_n + \check u_n + \mathcal{O}(L^{-1}),\\
	\hat u_n &= \Re\frac{-2}{\pi}\int_{0}^\infty  u_n(\tanh\pi w)\Re \psi(\frac{1}{2}+\mathrm{i} w)\mathrm{d}w,\\
	\check u_n &=\log\sqrt{2}+\sum_{i=1}^{\infty}\log (1+(4L)^{-2(2i)/n}X_{\frac{2i}{2n}}).
\end{align}
Here, the logarithmic term and $\hat u_n$ come from the $m=0$ term 
and $\check u_n$ come from the terms with $m=\pm1$  in Eq.~\eqref{gFH}.
To compute $\check u_n$, we perform a partial integral as in \cite{Calabrese:2010aab},
and pick up the poles of $\mathrm{d}u_n(\tanh\pi w)/\mathrm{d}w
=n(\coth n\pi w-\tanh \pi w)$. The poles of $\coth$ at $w=\pm \sqrt{-1}i$ with $i\neq 0$ combine to give the factor $\log(1+(4L)^{-2(2i)/n}X_{{i}/{n}})$, and the pole at $w=0$  give $(\log 2)/2$.

Plugging this into Eq.~\eqref{eq:fixedrho}  
we find that Eq.~\eqref{eq:scaling} holds with
\begin{equation}
Z^\text{conf}(q;1,\pm)=\frac12 \bigl[
	q^{\frac{1}{24}}\prod_{i=1}^\infty (1+q^{-(2i-1)})
\pm \sqrt{2}
q^{-\frac{1}{12}}\prod_{i=1}^\infty (1+q^{-2i/n})
\bigr],
\end{equation} which is $\chi_1(q^2)/2 + \chi_\varepsilon(q^2)/2 \pm \chi_\sigma(q^2)/\sqrt{2}$. The sign $\pm$ corresponds to $Z^\text{conf}(q;1,1)$ or  $Z^\text{conf}(q;1,\varepsilon)$, depending on the choice of projections in Eq.~\eqref{eq:projection}.

Finally, let us consider what happens when we use the clear-cut operation on one entangling point and the projection $(1+\sigma^z)/2$ on the other entangling point. In this case, 
the observables are generated by $2L+1$ gamma matrices $\gamma_{P,i}$ and $\gamma_{Q,i}$, acting on the Hilbert space of dimension $2^{L}$.
To obtain the reduced density matrix, note that the expectation value of a polynomial  $\mathcal{O}$  of these gamma matrices is $\braket{\mathcal{O}(1+\sigma^z)/2}$, but the term involving  $\sigma^z$ simply vanishes. Then 
the R\'enyi entropy is given by Eq.~\eqref{contourintegral} where $D_L$ is replaced by 
$D_L^\text{free-fixed}(\lambda)$ satisfying
$D_L^\text{free-fixed}(\lambda)^2 = D^{XX}_{2L+1}(\lambda)/\lambda^2$. Using the asymptotic form of $D^{XX}_L(\lambda)$ again, we can read off
\begin{equation}
	Z^\text{conf}(q;\sigma,1) =\frac1{\sqrt{2}} q^{\frac{1}{24}}\prod_i (1-q^{-(2i-1)})
\end{equation} from Eq.~\eqref{eq:scaling}. This reproduces $\braket{\sigma|q^{L_0+\bar L_0-c/12}|1}=(\chi_1(q^2)-\chi_\varepsilon(q^2))/\sqrt{2}$.

\section{Future directions}\label{sec:conclusions}

 Let us close the note by discussing some future directions the readers should pursue.
 First, the generalization to the entanglement entropies of two or more intervals and the mutual informations would be worthwhile.
 Second, our proposal applies in principle to QFTs in any dimensions, and it would be interesting to carry out explicit computations in higher dimensions.  
 Third, in the holographic analysis of the entanglement entropy as in \cite{Ryu:2006bv}, the fact that there is no tensor-product decomposition such as Eq.~\eqref{naive} is often neglected. This does not affect the leading piece of the entropy, but the issues discussed in this note might become non-negligible when one is interested in the subleading part of the holographic entanglement entropy, in particular if one wants to use it to better understand the gravity side. Also, a clear-cut tensor product decomposition like Eq.~\eqref{naive} of the total system is often assumed in the study of the black hole firewall paradox; the absence of such decomposition might play a role in the resolution.
 
\newpage
 
 \section*{Acknowledgments}
 It is a pleasure for the authors to thank helpful discussions with H. Casini, M. Honda, M. Huerta, H. Katsura, J. Maldacena, T. Nishioka, H. Ooguri, and K. Yonekura.
 The authors would also like to thank the anonymous referees for extremely fruitful feedbacks. 
This research was done  while the authors are visiting the Institute for Advanced Study, and the authors thank the hospitality there. 
KO is partially supported by the Programs for Leading Graduate Schools, MEXT, Japan,
%via the Advanced Leading Graduate Course for Photon Science 
and by JSPS Research Fellowship for Young Scientists.
YT is  supported in part by JSPS Grant-in-Aid for Scientific Research No. 25870159,
and in part by WPI Initiative, MEXT, Japan at IPMU, the University of Tokyo.

%\small\baselineskip=.9\baselineskip
%\let\bbb\bibitem\def\bibitem{\itemsep1pt\bbb}
\bibliographystyle{ytphys}
\bibliography{ref}

\end{document}